\documentclass[aps,twocolumn,showpacs]{revtex4}
\usepackage{graphicx}

\begin{document}

\title{Circular, elliptic and oval billiards in a
gravitational field}

\author{$^{1,2}$Diogo Ricardo da Costa, $^2$Carl P. Dettmann and $^{3,4}$Edson
D.\ Leonel}
\affiliation{$^1$Instituto de F\'isica da USP, Rua do Mat\~ao, Travessa R 187 
- Cidade Universit\'aria - 05314-970 - S\~ao Paulo - SP - Brazil\\
$^2$School of Mathematics, University of Bristol, Bristol, United Kingdom\\
$^3$Departamento de F\'isica, UNESP - Univ Estadual Paulista Av.24A, 1515 - 
13506-900 - Rio Claro - SP - Brazil\\
$^{4}$ The Abdus Salam - ICTP, Strada Costiera, 11 - 34151 - Trieste - Italy}

\date{\today} \widetext


\begin{abstract}

We consider classical dynamical properties of a particle in a constant gravitational
force and making specular reflections with circular, elliptic or oval boundaries.
The model and collision map are described and a detailed study of the energy
regimes is made.  The linear stability of fixed points is studied, yielding exact
analytical expressions for parameter values at which a period-doubling bifurcation
occurs. The dynamics is apparently ergodic at certain energies in all three models,
in contrast to the regularity of the circular and elliptic billiard dynamics in the field-free
case.  This finding is confirmed using a sensitive test involving Lyapunov weighted
dynamics.  In the last part of the paper a time dependence is introduced in the billiard
boundary, where it is shown that for the circular billiard the average velocity
saturates for zero gravitational force but in the presence of gravitational it
increases with a very slow growth rate, which may be explained using Arnold
diffusion. For the oval billiard, where chaos is present in the static case, the
particle has an unlimited velocity growth with an exponent of approximately $1/6$.
\end{abstract}
\maketitle

\section{Introduction}
\label{sec1}

In the 1920s billiards were introduced by Birkhoff \cite{ref1} into the theory
of dynamical systems. They consist of a point particle moving freely in a region
except for specular collisions with the boundary. Birkhoff's idea was to have a simple
class of models which shows the complicated behavior of non-integrable smooth
Hamiltonian systems without the need to integrate a differential equation \cite{ref1,ref2}.
Depending on the shape of the boundary, billiard dynamics may be (i) regular, with only
periodic or quasi-periodic orbits present; (ii) mixed, in which chaos, KAM islands (also
called periodic islands) and invariant spanning curves that limit the chaos in the systems
are present; (iii) completely ergodic, presenting only chaos in the phase space.

In recent years billiards continue to provide useful models for Hamiltonian dynamics, as
well as problems involving free motion in cavities and in extended structures.  The latter
includes the study of anomalous diffusion in geometries with infinite horizon \cite{ref3,ref_new_1,ref_new_2}.   Mixed phase space models include mushroom
billiards~\cite{ref4,extra_1}.  Time irreversible billiards have also been considered \cite{ref5}
as well as connections with many wave and quantum mechanical problems \cite{ref6,ref7,add2,add3}.  Further examples are reviewed in \cite{add0}.

It can be useful to open the billiard geometry by including a hole and investigating escape
time and related distributions.  Open billiards provide a useful starting point for an
understanding of more general classes of open dynamical systems \cite{add0} and includes
the study of open circular billiards and the Riemann hypothesis \cite{add1}.
Sometimes it is useful to investigate the survival probability or the
histogram of particles that reached certain height in the phase space
\cite{extra_diogo}.

An important subject in this area is the study of Fermi acceleration (FA)
\cite{ref8}. The phenomenon is characterized by an unlimited energy growth of
a bouncing particle undergoing collisions with a periodically moving and heavy
wall. According with the Loskutov-Ryabov-Akinshin (LRA) conjecture
\cite{ref9}, the introduction of a time dependence to the boundary of a
billiard is a sufficient condition to observe FA when the corresponding
static billiard has chaotic components. The elliptical case however must be
treated separately and the LRA conjecture does not apply for it. For the static
boundary it is integrable and hence has a phase space showing only regular
structures. There are two quantities which are preserved in the
elliptical billiard with static boundary: (i) energy ($E$) and; (ii) product of the angular
momenta about the two foci ($F$) as discussed in \cite{berry}. However, it was
shown recently \cite{ref10} that the introduction of a periodically time
perturbation to the boundary does lead to FA. The explanation for observing
diffusion in velocity is mainly related to the existence of a separatrix in
the phase space. According to \cite{ref10}, after introducing a time
perturbation to the boundary, the separatrix turns into a stochastic layer
yielding a type of turbulent behavior in $F$ therefore leading to diffusion in
velocity, hence producing the FA. More simulations were done in the model
\cite{diego_robnik} which confirmed the FA. Due to this observation in the
elliptical billiard and considering the LRA conjecture, a recent work
\cite{ref11} argues that the existence of heteroclinic fixed point in the phase
space may extend the conjecture in the
absence of chaos in the phase space. The circular billiard does not have such
a fixed point and remains regular even with vibrating boundaries.
Even in \cite{ref11}, the authors claim
that FA seems not to be a robust phenomena. The reflection law may be
modified so that the particle experiences a slightly inelastic collision
therefore having a fractional loss of energy upon collision. Even in such a
small limit of dissipation, the unlimited energy growth is suppressed.

It is also interesting to study billiards in the presence of a constant
gravitational field. The Galton board (1873) is a mechanical device that exhibits
stochastic behavior. It consists of a vertical (or inclined) board with interleaved
rows of pegs, where a ball moving into the Galton board moves under gravitation
and bounces off the pegs on its way down \cite{new_1}.  There have been many
recent investigations in gravitational billiards including a physical
experiment observing stable islands in chaotic atom-optics billiards
\cite{ref12}, characterization of the dynamics of a dissipative, inelastic
gravitational billiard \cite{ref13}, the study of linear stability in
billiards with potential \cite{ref2} and many others
\cite{ref15,ref16,ref17,livorati}. It also includes the wedge billiard
\cite{a1,a2,a3,ref13} which has singular regions in the phase space that
cannot be described by the KAM theorem \cite{ref16}.

The purpose of the present paper is to investigate some gravitational convex billiards,
with or without vibrating boundaries.  As noted above, the static and vibrating circular
billiards and static elliptical billiards have regular dynamics in the absence of a
gravitational field, while oval and vibrating elliptical billiards usually have mixed phase
space.  We apply the approach in Re. \cite{ref2} to characterise linear stability of
fixed points in the presence of the external field exactly in terms of the curvature of
the boundary and normal component of the velocity at the point(s) of collision.
A numerical search identifies apparently ergodic energy values, even for the circular
billiard; we then apply a sensitive Lyapunov weighted dynamics test to further confirm
this \cite{ref18}.  Ergodicity is interesting as it may suggest the existence of a new class
of ergodic billiards, extending known results for dispersing and defocusing non-gravitational
billiards on one hand and for the gravitational but piecewise linear wedge billiard on the
other.  In non-gravitational billiards it is known that no smooth convex billiard can be
ergodic~\cite{Lazutkin,Grigo}.

Finally, we introduce a time-dependence in the billiard, and show for the circle
in the absence of gravitational field and as expected, the average
velocity of the system approaches a regime of saturation. In the presence of
gravitational field, the velocity keeps growing for long number of collisions
but with a small slope of growth, an interesting effect since in the high velocity
limit the gravitational field has less and less effect, so the dynamics approaches
the non-gravitational regular behaviour. We can explain the continued but slow
acceleration in terms of Arnold diffusion.  For the breathing oval billiard, our
numerical result for the slope of growth is slightly larger than the one
obtained in the \cite{ref19} and theoretically foreseen in \cite{robnik}
but still of the same order of magnitude. 

This paper is organized as follows: In the section \ref{sec2} we describe
the model and obtain the mapping that describes the dynamical of a
particle. In section \ref{sec3} we study the energy regimes and explore the
phase space for different values of the control parameters. Section
\ref{sec4} is devoted to study some periodic orbits, for the oval, ellipse and
consider the low energy regime. Apparent ergodicity is studied in the section
\ref{sec5}. In section \ref{sec6} we take into account the time-dependent
billiards. Our conclusions and final remarks are presented in \ref{sec7}.

\section{The model and the map}
\label{sec2}

\begin{figure}[htb]
\includegraphics[width=1.0\linewidth]{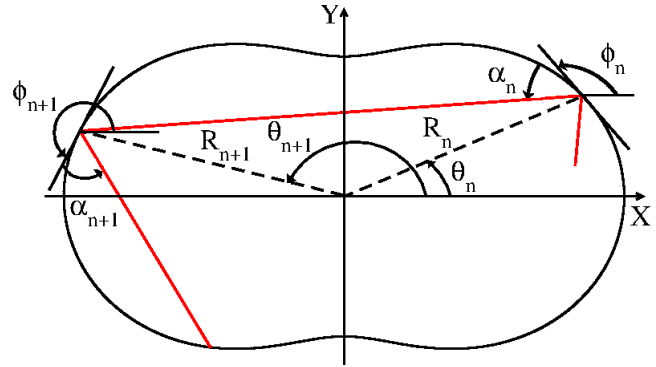}
\caption{Sketch of the boundary and angles considering $p=2$ and $\epsilon=0.3$ 
(oval billiard). As one can see we have locally negative curvatures in $\theta=\pi/2$ 
and $\theta=3\pi/2$. }
\label{Fig1_new}
\end{figure}

The models we are considering consist of a classical particle (or an ensemble
of non-interacting particles) confined inside and experiencing collisions with
a closed boundary of circular, elliptic and oval shapes under the presence of a
gravitational force. To describe the dynamics, we follow the same general
procedure as made in \cite{ref19}. Then the dynamics of each particle is
described in terms of a four-dimensional nonlinear mapping
$T(\theta_n,\alpha_n,\mid \overrightarrow{V}_n \mid,t_n)=
(\theta_{n+1},\alpha_{n+1},\mid \overrightarrow{V}_{n+1} \mid,t_{n+1})$  where
the variables denote: ($\theta$) the angular position of the particle;
($\alpha$) the angle that the trajectory of the particle forms with the
tangent line at the position of the collision; ($\mid \overrightarrow{V}
\mid$) the absolute velocity of the particle and; ($t$) the instant of the
collision with the boundary. The shape of the boundary is defined by its
corresponding radius in polar coordinates which is given by
\begin{equation}
R(\theta,p,\epsilon)=1+\epsilon\cos(p\theta)~,
\label{raio_ovoide} 
\end{equation}
where $\epsilon\in[0,1)$ is a parameter which controls the deformation of the
boundary (see for example \cite{berry}) and $p$ is a positive integer number.
For $\epsilon=0$ the circular billiard is recovered. 
In the Fig. \ref{Fig1_new} we have a sketch of the boundary and angles considering 
$p=2$ and $\epsilon=0.3$ (oval billiard). As one can see we have locally negative 
curvatures in $\theta=\pi/2$ and $\theta=3\pi/2$.
The radius of the elliptic
billiard is given by the following expression (see also
\cite{ref10,diego_robnik,ref11})
\begin{equation}
R(\theta,a,b)={ab\over
\sqrt{\left(b\cos\theta\right)^2+\left(a\sin\theta\right)^2}},
\label{raio_elipse}
\end{equation}
where $a$ and $b$ are one-half of the ellipse's major and minor axes. If
$a=b>0$ the circular billiard is observed. 

The rectangular components of the boundary at position $(\theta_n)$ are given
by
\begin{eqnarray}
X(\theta_n)&=&R(\theta_n)\cos(\theta_n)~, \\
Y(\theta_n)&=&R(\theta_n)\sin(\theta_n)~.
\label{eq402otd}
\end{eqnarray}
Starting with an initial condition $(\theta_n,\alpha_n,|{V}_n|,t_n)$, the
angle between the tangent at the boundary and the horizontal axis at the point
$X(\theta_n)$ and $Y(\theta_n)$ is given by
$\phi_n=\arctan[Y^{\prime}(\theta_n)/X^{\prime}(\theta_n)]$, where
\begin{equation}
X^{\prime}=dX/d\theta ~~~~\rm{and}~~~~ Y^{\prime}=dY/d\theta.  
\label{xlinha_ylinha}
\end{equation}

Since there is a gravitational force acting on the particle during the flight,
its trajectory is described by arcs of parabolas. For $t>t_n$ the position of
the particle as a function of time is given by
\begin{eqnarray}
X_p(t)&=&X(\theta_n)+|{\overrightarrow{V}}_n|\cos(\alpha_n+\phi_n)(t-t_n)~,\\
Y_p(t)&=&Y(\theta_n)+|{\overrightarrow{V}}_n|\sin(\alpha_n+\phi_n)(t-t_n)-{g(t-t_n)^2\over2}~.
\label{eq402otda}
\end{eqnarray}
Once the position of the particle as a function of the time is known, its
distance measured with respect to the origin of the coordinate system is given
by $R_{p}(\theta_p,t)=\sqrt{X^2_{p}(t)+Y^2_{p}(t)}$ and $\theta_{p}$ at
$(X_{p}(t), Y_{p}(t))$ is $\theta_{p}=\arctan[Y_{p}(t)/X_{p}(t)]$. The angular
position at the next collision of the particle with the boundary, i.e.
$\theta_{n+1}$, is numerically obtained by solving the following equation
$R(\theta,t)=R_p(\theta,t)$. The time is obtained by
\begin{eqnarray}
t_{n+1}=t_n+ {{\sqrt{\Delta X^2+\Delta Y^2}} \over
\vert\overrightarrow{V}_n\vert}~,
\label{eq06otd}
\end{eqnarray}
where $\Delta X=X(\theta_{n+1})-X(\theta_n)$ and $\Delta
Y=Y(\theta_{n+1})-Y(\theta_n)$. At the instant of collision, we use the
following reflection law
\begin{eqnarray}
{\overrightarrow{
V}}^{\prime}_{n+1}\cdot{\overrightarrow{T}}_{n+1}={\overrightarrow{V}}^{
\prime}_n\cdot{ \overrightarrow{T}}_{n+1}~,\\
{\overrightarrow{V}}^{\prime}_{n+1}\cdot{\overrightarrow{N}}_{n+1}=-{
\overrightarrow{V}}^{\prime}_n\cdot{\overrightarrow{N}}_{n+1}, 
\end{eqnarray}
where ${\overrightarrow{T}}$ and ${\overrightarrow{N}}$ are the unit tangent
and normal vectors and ${\overrightarrow{V}}^{\prime}$ denotes the velocity
of the particle measured in the moving referential frame. Based on these two
last equations, the particle's velocity components after collisions are
given by
\begin{eqnarray}
{\overrightarrow{V}}_{n+1}\cdot{\overrightarrow{T}}_{n+1}\nonumber
&=&V_x\cos(\phi_{n+1})+V_y\sin(\phi_{n+1})~, \\
{\overrightarrow{V}}_{n+1}\cdot{\overrightarrow{N}}_{n+1} &=&
V_x\sin(\phi_{n+1})-V_y\cos(\phi_{n+1}), \nonumber
\end{eqnarray}
where 
\begin{eqnarray}
V_x&=&|{\overrightarrow{V}}_n|\cos(\alpha_n+\phi_n),\nonumber \\
V_y&=&|{\overrightarrow{V}}_n|\sin(\alpha_n+\phi_n)-g(t_{n+1}-t_n)~. \nonumber
\end{eqnarray}

Therefore, the velocity of the particle at collision $(n+1)$ is given by
\begin{equation}
|\overrightarrow{V}_{n+1}|=\sqrt{({\overrightarrow{V}}_{n+1}\cdot{
\overrightarrow{
T}}_{n+1})^2+({\overrightarrow{V}}_{n+1}\cdot{\overrightarrow{
N}}_{n+1})^2}~. 
\end{equation}
Finally, $\alpha_{n+1}$ is obtained as
\begin{equation}
\alpha_{n+1}=\arctan\left[{{\overrightarrow{ V}}_{n+1}\cdot{\overrightarrow{
N}}_{n+1} \over {\overrightarrow{ V}}_{n+1}\cdot{\overrightarrow{ T}}_{n+1}
}\right].
\end{equation}

\section{Energy regimes}
\label{sec3}

\begin{figure*}[htb]
\includegraphics[width=1.0\linewidth]{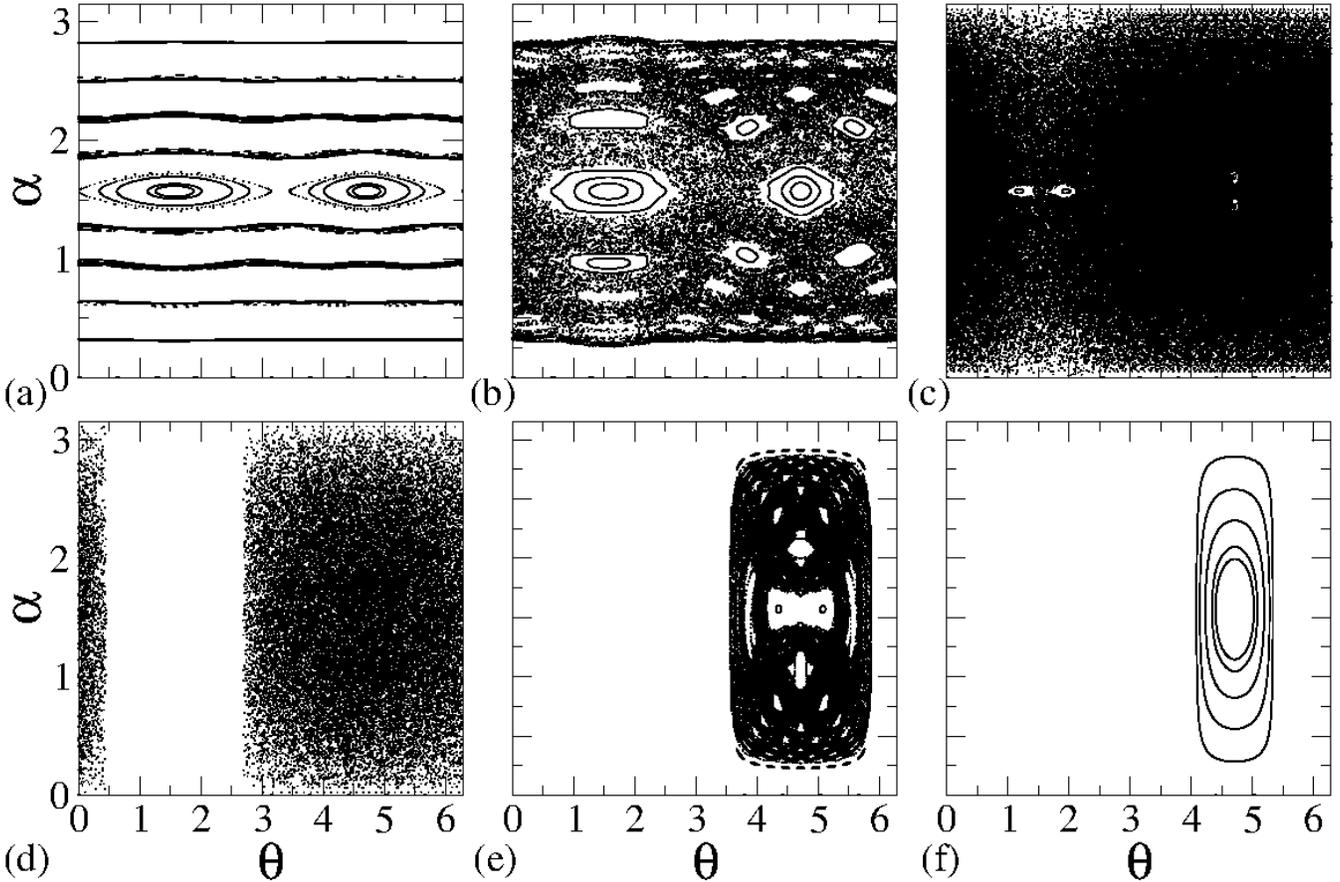}
\caption{Phase space for the circle ($\epsilon=0$) with $g=0.5$, considering:
(a) $E=2$; (b) $E=1.416$; (c) $E=1.088$; (d) $E=0.72$; (e) $E=0.3$; 
(f) $E=0.1$. }
\label{Fig1}
\end{figure*}
\begin{figure*}[htb]
\includegraphics[width=1.0\linewidth]{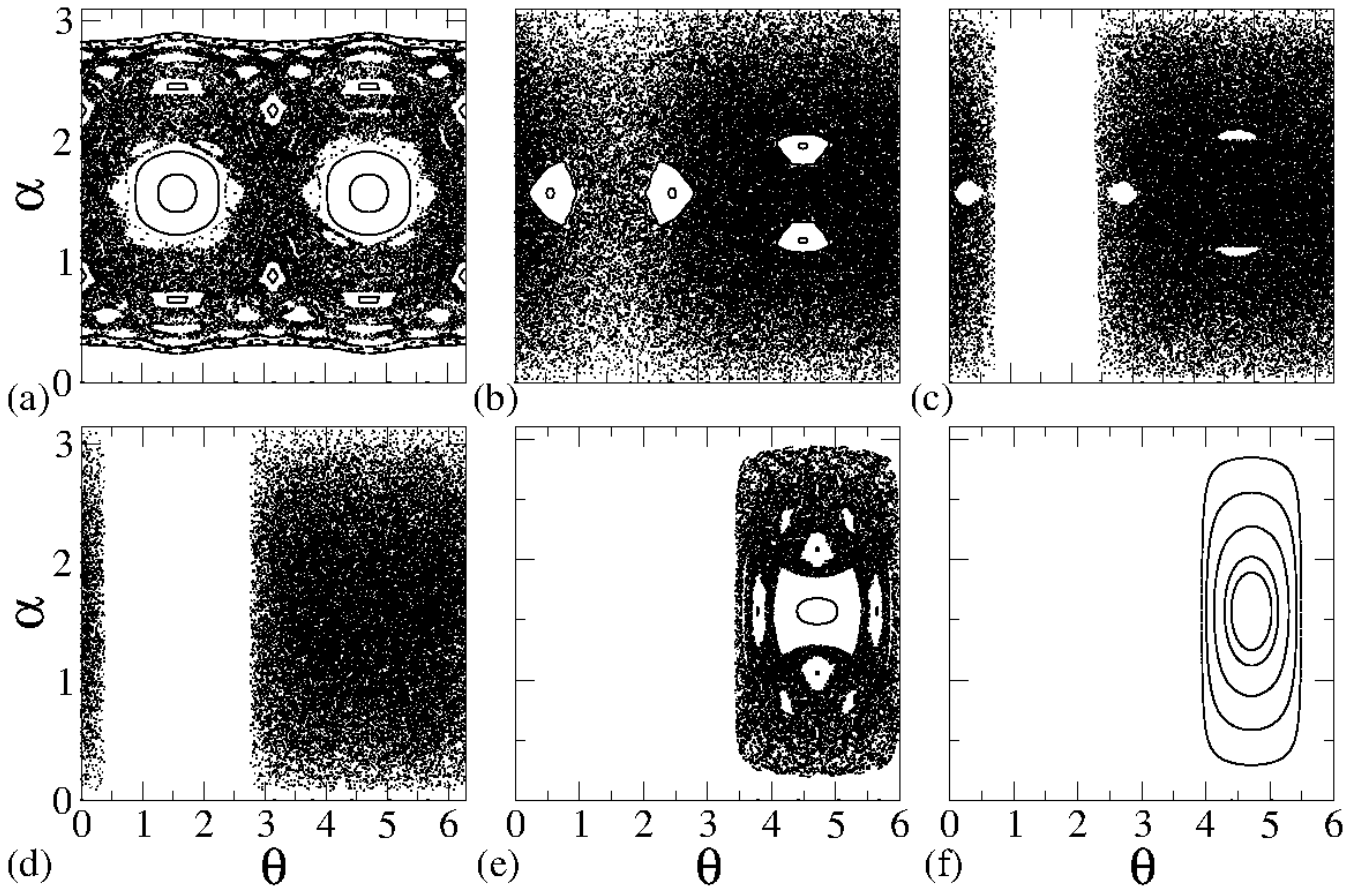}
\caption{Phase space for a convex oval billiard ($p=2$ and $\epsilon=0.1$). In
(a) $g=0$. For $g=0.5$ we have: (b) $E=1$; (c) $E=0.8$; (d) $E=0.651256$; (e)
$E=0.3$; (f) $E=0.1$.}
\label{Fig2}
\end{figure*}
\begin{figure*}[htb]
\includegraphics[width=1.0\linewidth]{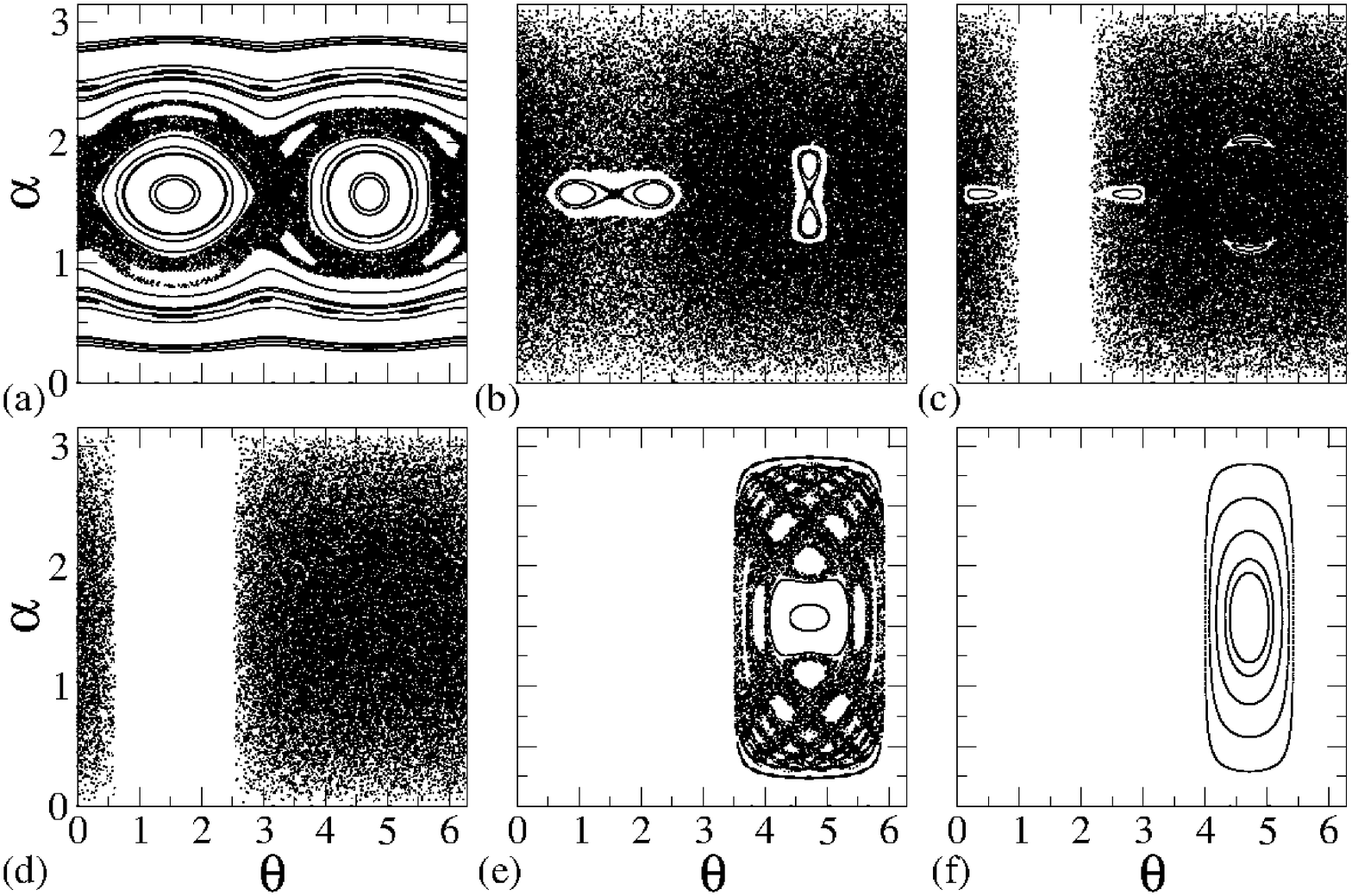}
\caption{Phase space for an elliptic billiard ($a=1.2$, $b=1.0$ and $g=0.5$)
considering: (a) $E=4$; (b) $E=1.28$; (c) $E=0.94$; (d) $E=0.824$; (e)
$E=0.3$; (f) $E=0.1$.}
\label{Fig3new}
\end{figure*}

In this section we present our numerical results for the circular, elliptic
and oval billiards. To construct the phase space of the systems, it is
essential to note the conserved energy $E=K+U$, where
$K=mV^2/2$ is the kinetic energy of the particle.
The potential energy is given by $U=mgh$, where $h$ is the
height of the particle measured with respect to an arbitrary reference
and $m$ is the mass of the particle considered as unitary in our simulations.
Equation (\ref{raio_ovoide}) has as its origin the center of the circle/oval
billiard, but we consider as a reference level the bottom of the billiard.
Therefore the reference level for the gravitational energy is obtained
applying $Y\rightarrow Y+Y_{ref}$, where $Y_{ref}$ is used to translate the
vertical axis and is given by
$Y_{ref}=R(\theta=\pi/2)\sin(p\pi/2)=1+\epsilon\cos(p\pi/2)$. 
Hence the potential energy of a particle is given by $U=g(Y(\theta)+Y_{ref})$,
leading to
\begin{equation}
U(p,\epsilon,\theta,g)=g\sin(\theta)[1+\epsilon\cos(p\theta)]
+1+\epsilon\cos(p\pi/2).
\label{u_1}
\end{equation}
For the elliptic billiard we have that $Y_{ref}=R(\theta=\pi/2)=b$. Setting
the energy constant, we write the initial velocity as
\begin{equation}
V_0=\sqrt{2[E-U(\theta_0)]}.
\label{eq_velocidade}
\end{equation}
If $E=0$ the particle does not have enough energy to leave the bottom of the
billiard, then we consider $E>0$.

Having constructed the collision map and incorporated energy conservation,
let us now investigate the dynamics looking at the phase space. We start
with the circle billiard, which can be obtained considering $\epsilon=0$ in
Eq. \ref{raio_ovoide} or $a=b>0$ in Eq. \ref{raio_elipse}. Figure
\ref{Fig1}(a) shows a phase space for the circular billiard considering the
energy $E=2$ and $g=0.5$. 

Each of the phase space shown in the figure was constructed considering a grid
of $10$ by $10$ equally spaced initial conditions in the intervals
$\theta_0\in[0,2\pi)$ and $\alpha_0\in[0,\pi)$, where the initial velocity is
given by Eq. (\ref{eq_velocidade}) and $t_0=0$. Each initial condition
was iterated up to $2000$ collisions with the boundary. For $E=2$ and $g=0.5$,
as shown in Fig. \ref{Fig1}(a), we see two islands around
$\theta=\pi/2$ and $\theta=3\pi/2$ (for $\alpha=\pi/2$), defining a period
two periodic orbit in the center. In the limit of $E\rightarrow\infty$ the
velocity of the particle increases and the arcs of parabola describing the
motion of the particle become straight lines. The phase space for the circular
billiard in the absence of gravitational field billiard is recovered, and only
periodic and quasi-periodic obits are observed.

Decreasing the energy to $E=1.416$ as shown in Fig. \ref{Fig1}(b) we see the
phase space is of mixed type. Now changing the energy to $E=1.088$, we see the
fixed points in $\theta=\pi/2$ or $\theta=3\pi/2$ become unstable, creating a
period four elliptic periodic orbit. For $E=0.72$ (see Fig. \ref{Fig1}(d)) KAM
islands are not observed and an apparent ergodic region emerges. More details
are shown later. 

Decreasing the energy to $E=0.3$ (see Fig. \ref{Fig1}(e)) the chaotic sea
occupies a small region of the phase space and some KAM islands are now
visible. Decreasing yet more the energy to $E=0.1$ (Fig. \ref{Fig1}(f)) we
see that periodic (or quasi periodic) orbits are present and chaos is no
longer present. The fixed point in $\theta=3\pi/2$ is stable (elliptical).
More details about the change of stability of the fixed points are 
shown in the section \ref{sec4}.

Let us study the phase space for a convex oval billiard considering
$\epsilon=0.1$ and $p=2$. In Fig. \ref{Fig2}(a) we see a phase space for
$g=0$ (absence of gravitational field). The phase space is then of mixed type.
Another way to obtain Fig. \ref{Fig2}(a) for $g=0.5$ is to consider the
energy $E\rightarrow\infty$. In Fig. \ref{Fig2}(b) $g=0.5$ for a constant
energy $E=1$ while fixed point in $\theta=\pi/2$ and $3\pi/2$ becomes unstable
and creates a period four elliptical fixed point. Considering $E=0.8$ and for
fixed value of $g=0.5$ we observe in Fig. \ref{Fig2}(c), that there is a
region where the orbits cannot access. This happens because of the
energy conservation and the particle does not have enough energy to reach such 
regions. Figure \ref{Fig2}(d), with $E=0.651256$,
shows a phase space where KAM islands are not observed. Decreasing the value
of energy to $E=0.3$ (see Fig. \ref{Fig2}(e)) the chaotic sea is confined in a
small region of the phase space. Changing the energy to $E=0.1$ (Fig.
\ref{Fig2}(f)) we see only periodic (or quasi periodic) orbits are observed
and the chaos is no longer present, as similar to Fig. \ref{Fig1}(a). 

Results for the elliptic billiard are shown in Fig. \ref{Fig3new}. The control
parameters used in the simulations were $a=1.2$, $b=1$ and $g=0.5$. Figure
\ref{Fig3new}(a) shows the phase space considering $E=4$. As one can see the 
chaotic sea is observed near $\alpha\cong\pi/2$, and periodic or quasi
periodic regions can be observed above and below such $\alpha$. For
$E\rightarrow\infty$ only periodic or quasi-periodic orbits are observed,
therefore the results for the absence of gravitational field are recovered.
Of course this case can also be obtained considering $g=0$. The fixed points
in $\theta=\pi/2$ and $\theta=3\pi/2$ (for $\alpha=\pi/2$) are elliptical
(stable). Decreasing the value of the energy for $E=1.288$ produces a phase
space where a saddle fixed point can be observed in $\theta=\pi/2$ and
$\theta=3\pi/2$, as shown in Fig. \ref{Fig3new}(b). Decreasing even more the
energy turns the fixed point unstable (hyperbolic fixed point). The phase
space for $E=0.94$ is shown in Fig. \ref{Fig3new}(c), and as one can see the
fixed point in $\theta=\pi/2$ is no longer available. Again there is a
forbidden region that appears due to the conservation of energy. For $E=0.824$
the phase space is apparently ergodic (see Fig. \ref{Fig3new}(d)). Figure
\ref{Fig3new}(e) shows the chaotic sea is observed in a small region of
the phase space, and the number of KAM islands is quite large. Finally, in the
limit $E\rightarrow0$ the phase space presents periodic and quasi-periodic
regions, as shown in Fig. \ref{Fig3new}(f) considering $E=0.1$, again
similar to the previous cases.

To conclude, we see that the phase space for the circle, elliptic and
oval billiard have some common characteristics: all of them present apparent
ergodic regions for some combinations of control parameters; for
$E\rightarrow0$ the chaotic sea tends to disappear; these systems present
changes of stability for the fixed points in $\theta=\pi/2$ or $\theta=3\pi/2$
($\alpha=\pi/2$). However, despite of the similarities pointed here, there are also 
marked differences.

\subsection{Supplemental Data I}

As attached files we have three different videos showing details about the
phase spaces of Fig. \ref{Fig1}, Fig. \ref{Fig2} and Fig. \ref{Fig3new}. For
all, we have considered $g=0.5$ and the energy is varied in the interval
$E\in(0,2)$. The phase space $\alpha ~\rm{vs}~ \theta$ for the circle, oval
and elliptic billiards can be found in the following addresses: \href{http://www.youtube.com/watch?v=Brb3GHXjfFE&feature=share&list=UUG_8BL4kcVV1rYYR4Zyn9ew}{http://www.youtube.com/watch?v=Brb3GHXjfFE&feature=c4-overview&list=UUG_8BL4kcVV1rYYR4Zyn9ew}, 
\href{http://www.youtube.com/watch?v=heWA0HPJcGQ&feature=share&list=UUG_8BL4kcVV1rYYR4Zyn9ew}{http://www.youtube.com/watch?v=heWA0HPJcGQ&feature=share&list=UUG_8BL4kcVV1rYYR4Zyn9ew} and
\href{http://www.youtube.com/watch?v=BM4oSeXTGxk&feature=c4-overview&list=UUG_8BL4kcVV1rYYR4Zyn9ew}{http://www.youtube.com/watch?v=BM4oSeXTGxk&feature=c4-overview&list=UUG_8BL4kcVV1rYYR4Zyn9ew}. 
To construct the frames, we used a grid of $500$ by $500$
different initial conditions and each orbit was iterated up to $2000$ times.
As we can see there exist a lot of duplication of periods and a complicated
behavior can be observed. 

\section{Periodic orbits}
\label{sec4}

\begin{figure}[htb]
\includegraphics[width=1.0\linewidth]{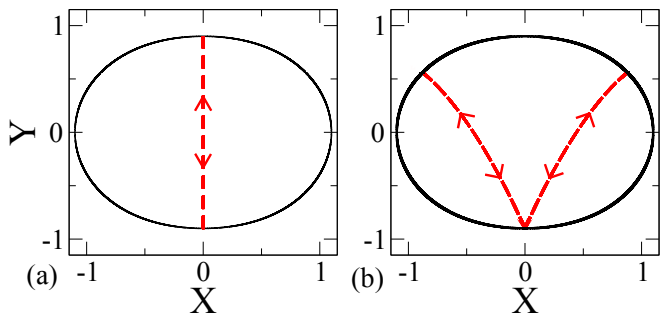}
\caption{(Colour online) For the oval billiard with $p=2$ and $\epsilon=0.1$
we have in (a) the trajectory for an elliptic fixed point in $\theta=\pi/2$ or
$\theta=3\pi/2$ ($\alpha=\pi/2$) when considering $g=0$. As one can see the
particle bounces between the top and bottom parts of the billiard. In (b), for
$g=0.5$ and $E=1$, we can an example of trajectory after the bifurcation, with
the creation of a period four fixed point.}
\label{Fig3}
\end{figure}

In this section we concentrate to study the linear stability of some periodic
orbits in the phase space. We start with investigating what happens with the 
fixed points $\theta=\pi/2$ and $\theta=3\pi/2$ (for $\alpha=\pi/2$) in Figs. 
\ref{Fig1}(b,c) or \ref{Fig2}(a,b). In them we see the left fixed point
($\theta=\pi/2$ and $\alpha=\pi/2$) in the Fig. \ref{Fig1}(b) and Fig.
\ref{Fig2}(a) is elliptical and after decreasing the energy (Figs.
\ref{Fig1}(c) and \ref{Fig2}(b)), it becomes unstable and a period four fixed
point arises. The duplications in the left occur in the horizontal axis and
in the right occurs in the vertical. It happens this way because the
trajectories are arcs of parabola for $g>0$ and as can be seen in Fig.
\ref{Fig3}(b) it needs to hit another part (by looking at two separate parts
measured with respect to an imaginary line cutting/separating the billiard at
$x=0$) of the billiard to continue in a periodic orbit. This behavior is
different for $g=0$ or $E\rightarrow\infty$, where the fixed point hits the
top and bottom of the billiard (see Fig. \ref{Fig3}(a)). Another way to
observe the phenomenon in Fig. \ref{Fig2}(a,b) is to consider initially $g=0$
and increasing it until the fixed point becomes unstable. 

\begin{figure}[htb]
\includegraphics[width=1.0\linewidth]{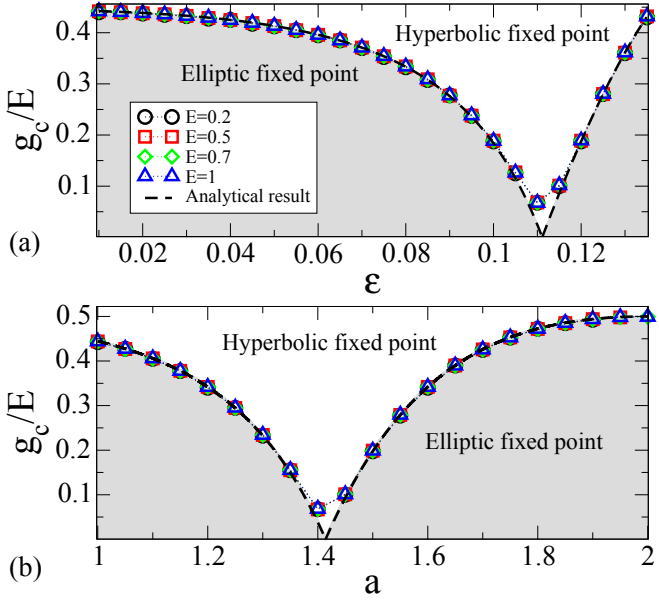}
\caption{(Colour online) For $t_0=0$, $\theta_0=\pi/2+10^{-8}$ and
$\alpha_0=\pi/2$ we have the numerical and analytical results for: (a)
$g_c/E~\rm{vs}~\epsilon\in[0.01,0.135]$ for the oval billiard with $p=2$ and
considering four different values of energy $E$; (b) $g_c/E~\rm{vs}~a\in[1,2]$
for the elliptic billiard with $b=1$.}
\label{Fig4}
\end{figure}

The position ($\theta$, $\alpha$) of the elliptical fixed points in
$\theta\cong\pi/2$ and $\theta\cong3\pi/2$ (for $\alpha=\pi/2$) can be
obtained numerically. Therefore it is possible to find the critical value of
$g$ in which there is a change from stable to unstable (saddle) fixed point
denoted as $g_c$. We then obtain a period-doubling bifurcation. For the oval
billiard, we have in Fig. \ref{Fig4}(a) $g_c/E$ as function of the control
parameter $\epsilon$ and using four different values of $E$. We rescaled the
vertical axis by $E$ because it is proportional to $g$. Then $g_c/E$ is a
constant as can be seen in Fig. \ref{Fig4}(a) looking at the fact that for
different values of $E$, the curve has an universal behavior. Therefore,
these curves of $g_c/E$ as function of $\epsilon$ are scaling invariant for
$E$. Under this curve, the fixed point in $\alpha=\pi/2$ and $\theta=\pi/2$
(or $\theta=3\pi/2$) is elliptical and above this, it is a unstable
(hyperbolic). 

Similar analysis can be done in the elliptic billiard. The results of $g_c/E$ 
as function of the control parameter $a$ for different values of energy are
shown in Fig. \ref{Fig4}(b) for $b=1$. The curves are again scaling invariant
for $E$.

The behavior of the fixed points can be obtained analyzing the linear
stability. To compute this it is convenient to introduce Green's residue $R$
and its complement $\bar{R}$, which are defined by
$
\bar{R}=1-R=\rm{tr}^{(4)}/4=(2+\rm{tr})/4,
$
where
$\rm{tr}^{(4)}_j=\rm{tr}\prod^j_{i=1}\textbf{M}^{(4)}_i\textbf{K}^{(4)}_i$. 
$\textbf{K}^{(4)}$ is the linearized reflection matrix. The stability is
split up into contributions from the reflection $\textbf{K}^{(4)}$ and from
the free motion $\textbf{M}^{(4)}$. $j$ is the period of the orbit. More
details can be obtained in the Ref. \cite{ref2}.

We use the following formulae to calculate the Residue of symmetric
orbits:
\begin{equation}
R=4R^{\prime}(1-R^{\prime})~~~~~ \bar{R}=\left(1-2\bar{R}^{\prime}\right)^2.
\label{r_and_rprime}
\end{equation}
For the fixed points mentioned in Fig. \ref{Fig2}(a), i.e., orbits that
touch the top and bottom of the billiard, we show the
residue for $\alpha=\pi/2$ and $\theta=\pi/2$ (or $\theta=3\pi/2$) is equal to
\cite{ref2}
\begin{equation}
 R^{\prime}=1-V^*\kappa\tau^{\prime},
 \label{eq_rprime}
\end{equation}
where $\tau^{\prime}=\tau/2$ and $\tau$ is the time spend for the particle to
return to the initial point. $V^*$ is the normal component of the velocity and
$\kappa$ is the curvature in the collision point. It is important to say that
if $0<R<1$ the fixed point is elliptical.

For $E>2g(1-\epsilon)$ we ensure the particle touches the top and bottom
of the billiard (see Fig. \ref{Fig3}(a)). The time $\tau^{\prime}$ for the
particle to travel the distance between the top and bottom (or vice versa) is
equal to 
\begin{equation}
 \tau^{\prime}={V_d-V_u\over g}.
\end{equation}
Here $V_d=\sqrt{2E}$ and $V_u=\sqrt{2E-4gR(\theta=\pi/2)}$ are the
velocities, respectively, in the bottom part (where the potential energy $U$
is equal to zero) and in the upper part of the billiard.

For billiards, the curvature $\kappa$ can be obtained using the following 
expression (see for example \cite{diego_cnsns})
\begin{equation}
\kappa(\theta)={X^{\prime}(\theta)Y^{\prime\prime}(\theta)-X^{\prime\prime}Y^{
\prime}(\theta)\over\left[X^{\prime2}(\theta)+Y^{\prime2}(\theta)\right]^{3/2}
},
\label{eq_curv}
\end{equation}
where $X^{\prime\prime}=d(X^{\prime})/d\theta$ and
$Y^{\prime\prime}=d(Y^{\prime})/d\theta$ ($X^{\prime}$ and $Y^{\prime}$ were
previously defined in Eq. (\ref{xlinha_ylinha}).

In order to better present the analytical and numerical results, we separate
the findings in two different parts. First we analyse the oval/circle
billiards and after that the results for the elliptic billiard are obtained. 

\subsection{Oval}

For the oval billiard, when considering $\theta=\pi/2$ or $\theta=3\pi/2$, the 
curvature, using the Eq. (\ref{eq_curv}), is written as
\begin{equation}
\kappa={R\left[R-{R^{\prime\prime}}^2\right]+2{R^{\prime}}^2
\over\left(R^2+{R^{\prime\prime}}^2\right)^{3/2}}.
\label{kappa_oval}
\end{equation}
In this case $R=R(\theta=\pi/2)$,
$R^{\prime}={\partial{R}\over\partial{\theta}} (\theta=\pi/2)$
and $R^{\prime\prime}={\partial{R^{\prime}}\over\partial{\theta}}(\theta=\pi/2)$.

After obtaining $R^{\prime}$ in Eq. (\ref{eq_rprime}), we can evaluate $R$ in
Eq. (\ref{r_and_rprime}). Considering $R=0$ and $V^*=V_u$, we show that $g_c$
is given by
\begin{equation}
g_c=-E{4\kappa\left[1+2\kappa(\epsilon-1)\right]
\over
\left[1+4\kappa(\epsilon-1)\right]^2},
\label{g_c_left}
\end{equation}
where $\kappa$ is given by Eq. (\ref{kappa_oval}). As one sees from previous 
equation, $E$ is scaling invariant. If we define a new variable $g/E$ the
result obtained for different $E$ are basically the same, confirming the 
rescaling made in Fig. \ref{Fig4}(a). 

Now considering $R=0$ and $V^*=V_d$, we found that $g_c/E$ is written as
\begin{equation}
{g_c\over E}=2\kappa-4\kappa^2(\epsilon-1)+2\kappa\left[1+2(\epsilon-1)\right].
\label{g_c_right}
\end{equation}

We need to find the condition in which Eq. (\ref{g_c_left}) and Eq.
(\ref{g_c_right}) have the same values, and the value of $g$ when it happens
is given by
\begin{equation}
g^*={1\over2}{(2\kappa-1)\over\kappa}.
\label{g*}
\end{equation}
According to our results, Eq. (\ref{g_c_left}) fits the numerical results for 
$g<g^*$, and Eq. (\ref{g_c_right}) is used when $g>g^*$. The results here are
general, applying to every control parameter $g$, $E$,
$\epsilon$ and $p$ of the oval billiard.

For a particular case, with $p=2$ (convex oval billiard) we have that the
curvature is equal to
\begin{equation}
\kappa(\epsilon)={(1-5\epsilon)\over\left(1-\epsilon\right)^2}.
\label{kappa_oval_particular_case}
\end{equation}
In this situation, we value of $g^*$ is equal to $1/9$ (see Eq. (\ref{g*})).
The results for $\epsilon<1/9$ in Fig. \ref{Fig4}(a) can be obtained
considering $R=0$ and $V^*=V_u$. As result we obtain analytically a curve for
$g_c$ as function of $\epsilon$ which is equal to
\begin{widetext}
\begin{equation}
g_c=E{(5\epsilon-1)\over(19\epsilon-3)(\epsilon-1)}\left[
{
a_1\left(a_2\epsilon-a_3+|1-9\epsilon|\sqrt{2}\right)
\over
19\epsilon-3
}
-2
\right],
\end{equation}
\end{widetext}
where $a_1=\sqrt{2}$, $a_2=10a_1$ and $a_3=a_2/5$. For $\epsilon=0$ (circle
billiard), the value of $g_c$ is equal to $4/9$. For $\epsilon>g^*=1/9$, it is
necessary to consider $V^*=V_d$ and $R=0$, and the result obtained is
\begin{equation}
g_c=E{(5\epsilon-1)\over(\epsilon-1)^2}\left[
{
a_1\left(a_2\epsilon-a_3+|1-9\epsilon|\sqrt{2}\right)
\over
\epsilon-1
}
-2
\right].
\end{equation}
As one can see for both cases dividing both sides of the equations for $E$
yields $g_c/E$ , which is a function of $\epsilon$, therefore confirming the
results shown in Fig. \ref{Fig4}(a). Moreover the critical value of $g$ is
scaling invariant for the energy $E$. The analytical results obtained can be
observed in Fig. \ref{Fig4}(a) as the black dashed curves. As one can see, the
numerical and analytical results are in good agreement. 

\subsection{Ellipse}

Considering the elliptic billiard with radius given by Eq.
(\ref{raio_elipse}), the curvature for $\theta=\pi/2$ or $\theta=3\pi/2$ is
given by
\begin{equation}
\kappa={b\over a^2}.
\label{kappa_elipse}
\end{equation}

First of all we solve $R=0$ considering $V^*=V_d$. After some calculations,
the result obtained is
\begin{equation}
{g\over E}=-{2b\over a^4}{\left(a^2-2b^2+\left|a^2-2b^2\right|\right)}.
\label{g_elipse_right}
\end{equation}
The solution of $R=0$ and considering $V^*=V_u$ is given by
\begin{equation}
{g\over E}={2b\left(2b^2-a^2+\left|a^2-2b^2\right|\right)\over
\left(a^2-4b^2\right)^2}.
\label{g_elipse_left}
\end{equation}
Combining Eq. (\ref{g_elipse_left}) and Eq. (\ref{g_elipse_right}) we have 
\begin{equation}
a=\sqrt{2}b.
\label{eq_critical_a}
\end{equation}
After analysing either the numerical and analytical results, we observe that
Eq. (\ref{g_elipse_right}) and Eq. (\ref{g_elipse_left}) are used for
$a>\sqrt{2}b$ and $a<\sqrt{2}b$, respectively. These analytical results are
confirmed in Fig. \ref{Fig4}(b) as the black dashed lines, where the
analytical and numerical results are in good agreement. For the situation
shown in Fig. \ref{Fig4}(b), with $b=1$, we see that the minimum value of
$g_c/E$ is $\sqrt{2}$ (see Eq. (\ref{eq_critical_a})). As known, for
$a=b=1$, we have a circle with unitary radius. Considering Eq.
(\ref{g_elipse_left}), we show that $g_c/E=4/9$. This is the same result
obtained when considering $\epsilon=0$ in the oval billiard.

It is interesting to see that the eccentricity of a ellipse is defined as
$\sqrt{1-\left(b/a\right)^2}$, therefore $g_c=0$ when the eccentricity is
$1/2$.

\begin{figure}[htb]
\includegraphics[width=1.0\linewidth]{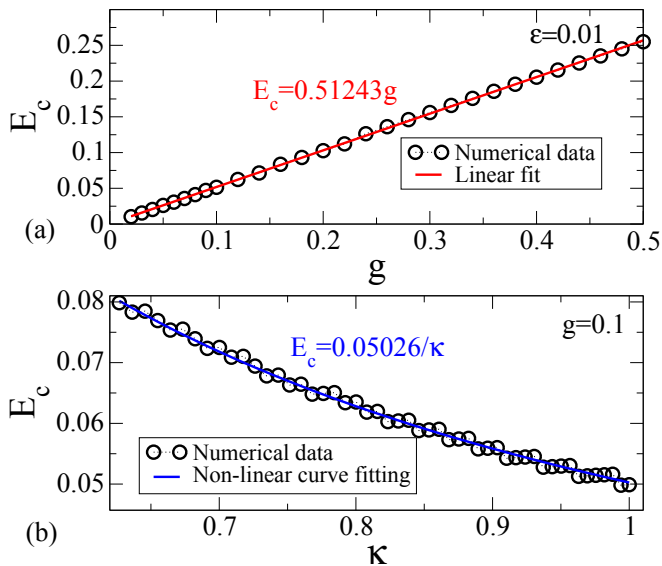}
\caption{(Colour online) For $p=2$, $t_0=0$, $\theta_0=3\pi/2$ and
$\alpha_0=\pi/2$ we have: (a) $E_c~\rm{vs}~g$ for $\epsilon=0.01$. After a
linear fit, the slope obtained is $0.51243$; (b) $E_c~\rm{vs}~\kappa$
considering $g=0.1$. A non-linear curve fitting produces as result
$E_c=0.05026/\kappa$.}
\label{Fig5}
\end{figure}

\begin{figure}[htb]
\includegraphics[width=1.0\linewidth]{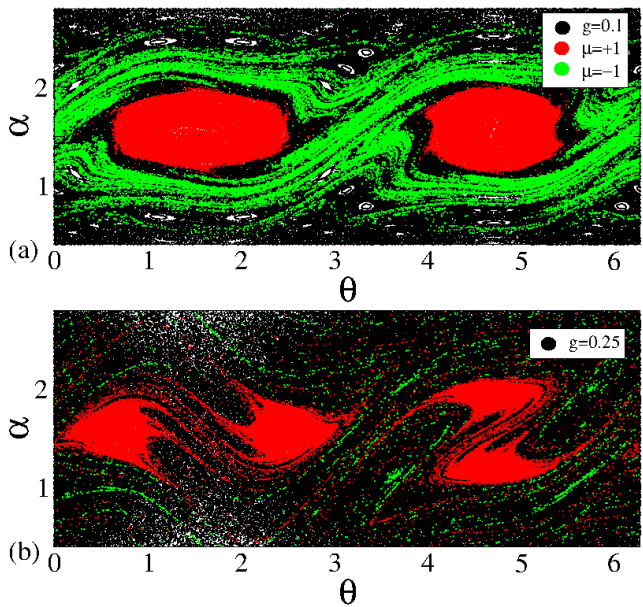}
\caption{(Colour online) Phase space (black dots) for different values of g: 
(a) $g=0.1$; (b) $g=0.25$. As red colour we have the LWD considering 
$\mu=+1$ and in green colour $\mu=-1$. We used $E=0.5$, $p=2$ and
$\epsilon=0.1$.}
\label{Fig6}
\end{figure}

\begin{figure}[htb]
\includegraphics[width=1.0\linewidth]{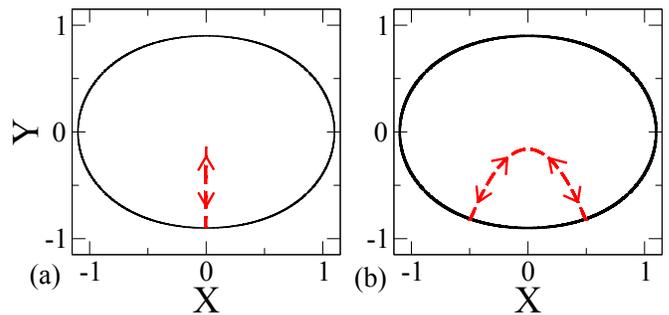}
\caption{(Colour online) An example of period one fixed point with $g=0.5$, $p=2$, 
$\epsilon=0.1$ and $E=0.38$. In (b) we have a bifurcation and the creation of a 
period two fixed point for $E=0.416$. As one can see, the orbits do not reach the
top for the billiard.}
\label{orbita2}
\end{figure}

\begin{figure*}[t]
\includegraphics[width=1.0\linewidth]{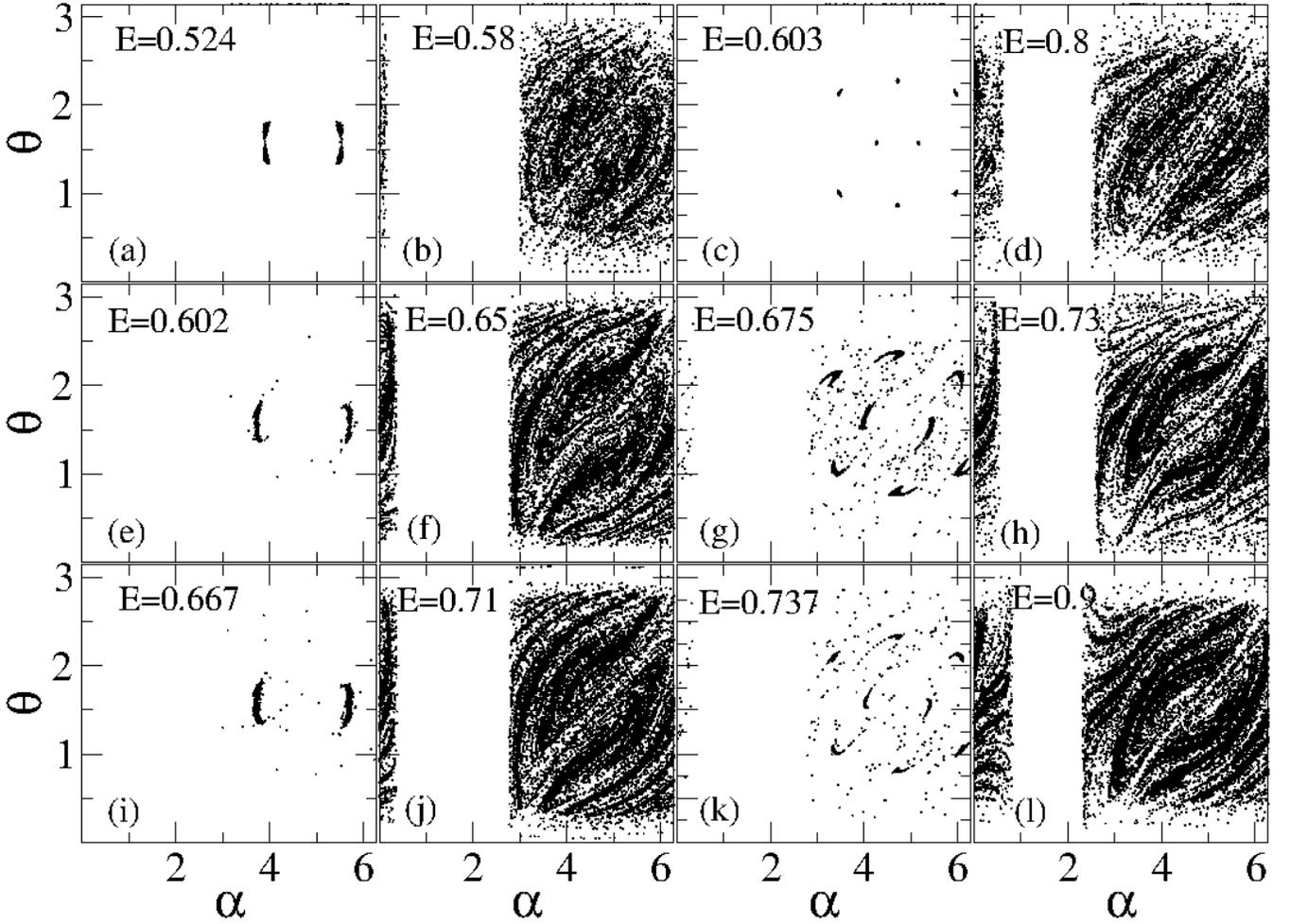}
\caption{For different values of $E$ we have the position of $10^{4}$ 
the trajectories and clones using the Lyapunov weighted 
dynamics (LWD) after $10^3$ iterations and for $\mu=-1$. 
We have considered $\epsilon=0$ (circle billiard) in (a), (b), (c) and (d).
In (e), (f), (g) and (h) the control parameters used were $p=2$, 
$\epsilon=0.1$ and $g=0.5$ for the oval billiard and in (i), (j), (k) and (l) we
have considered $a=1$ for the elliptic billiard.} 
\label{lwd_compara}
\end{figure*}
\begin{figure*}[t]
\includegraphics[width=1.0\linewidth]{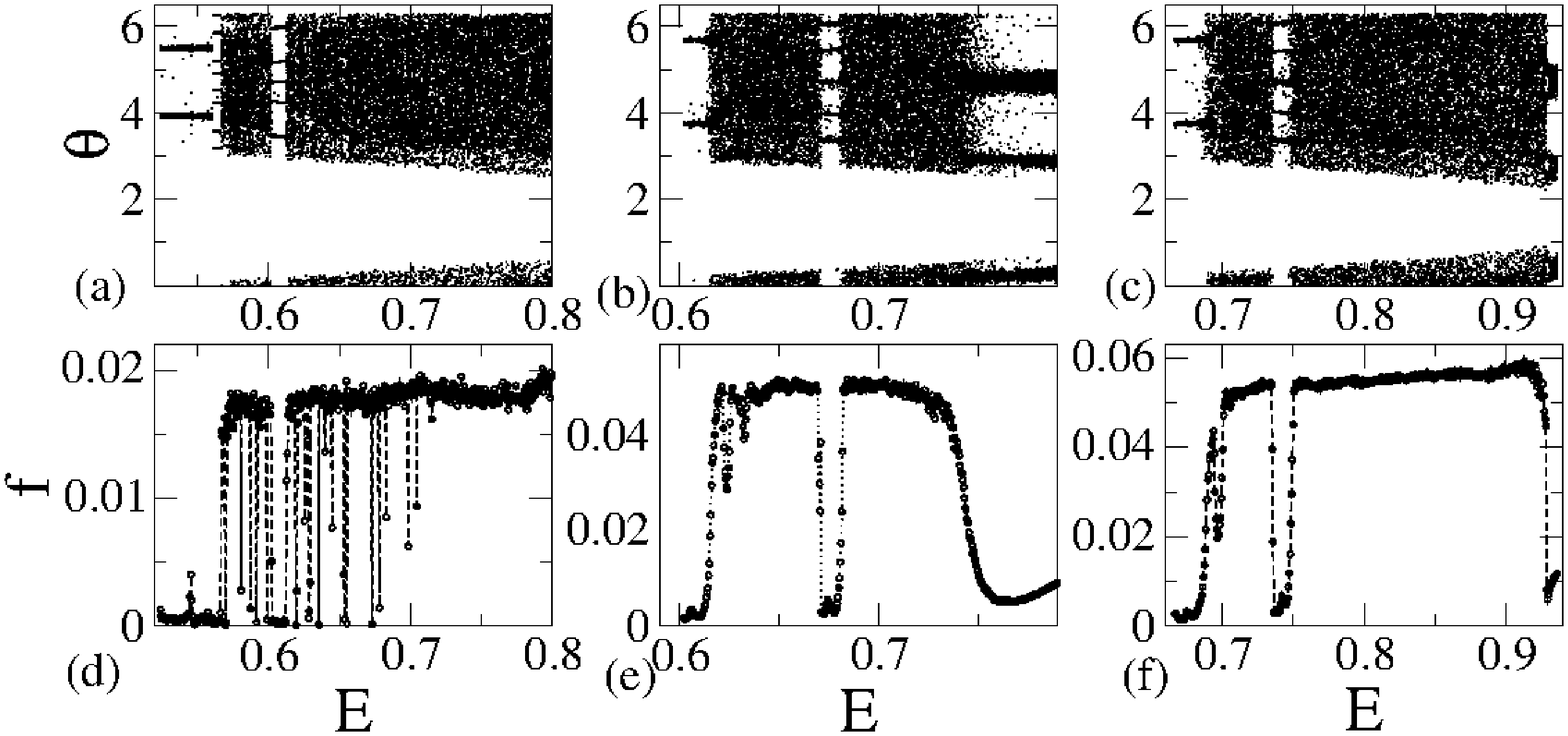}
\caption{For $g=0.5$ and $\mu=-1$ we have the 
$\theta$ coordinate for the LWD as function of $E$ and considering the circle in (a), 
oval ($p=2$ and $\epsilon=0.1$) in (b) and the ellipse ($b=1$ and $a=1.2$) in (c).
In (d), (e) and (f) we have the fraction $f$ for the number of occupied rectangles 
as function of $E$ for the circle, oval and elliptic billiard, respectively.} 
\label{Fig8new_2}
\end{figure*}

\subsection{Low energy}

The results previously described cannot be applied to orbits that start in
$\theta=3\pi/2$ ($\alpha=\pi/2$) and do not reach the top boundary of the
billiard. As example Fig. \ref{orbita2}(a) shows a period one orbit with
$g=0.5$, $p=2$, $\epsilon=0.1$ and $E=0.38$. It happens for orbits with
$E<2gR(\theta=\pi/2)$, where a period one fixed point is present. One can
observe that the fixed point in $(\theta,\alpha)=(3\pi/2,\pi/2)$ is elliptical
in Figs. \ref{Fig1}(f) and \ref{Fig2}(e,f), where $E$ is a small value.
Increasing $E$, the stable period one fixed point undergoes an inverse parabolic
transition creating an elliptic period two orbit in a period doubling
bifurcation, as shown in Figs. \ref{Fig1}(e) and \ref{Fig2}(d). Another
example of such a bifurcation is shown in Fig. \ref{orbita2}(b), for
$g=0.5$, $p=2$, $\sigma= 0.1$ and $E=0.416$, where a period two fixed point
does not touch the top of the billiard. It happens because for focusing
boundaries the orbit is elliptic for small energies and becomes inverse
hyperbolic if the energy becomes large compared with the radius of
curvature. 

For the oval billiard, a good method to detect the bifurcation is to follow
the position of the fixed points when varying $E$ from $0$ to
$2gR(\theta=\pi/2)$. We call $E_c$ as the energy in which the first
period-doubling bifurcation is observed. In Fig. \ref{Fig5}(a) we have $E_c$
as function of $g$ for $\epsilon=0.01$ and $p=2$, and after a linear fit we
observe that $E_c=0.51243g$ ($E_c\propto g$). $E_c$ can be obtained by
varying $\epsilon$(consequently the curvature $\kappa$, which is given by Eq.
(\ref{kappa_oval_particular_case})). It was done in Fig. \ref{Fig5}(b) using
$g=0.1$. After fitting with a non-linear curve we obtained
$E_c=0.05026/\kappa$ ($E_c\propto 1/\kappa$). We conclude that $E_c\propto
g/\kappa$. Such result can be proved analytically if one consider that
$\tau=2V^*/g$ for the residue $R=2\kappa E/g$. We then obtain
\begin{equation}
g=2E\kappa, 
\end{equation}
where $E$ is numerically equal to the kinetic energy for $\theta_0=3\pi/2$, 
i.e., $E={V^*}^2/2$.

Similar results can be found for the elliptic billiard, where one needs to
consider the curvature given by Eq. (\ref{kappa_elipse}).

\section{Ergodicity}
\label{sec5}

In this section we test the ergodicity of the dynamics appearing in the above phase
space plots, using a tool called Lyapunov weighted dynamics (LWD) to
identify rare physical trajectories in the phase space \cite{ref18}. Basically
the method consists of considering a number of $10^4$ trajectories 
equally distributed in the phase space. Each has a perturbed trajectory, separated
by $\delta\theta_i=10^{-3}$. Both are evolved in in phase space and they are 
perturbed by a weak random
force of intensity $\sqrt{\sigma}=10^{-6}$ that slightly modifies the angle
$\theta$. If the particle reaches the boundary of the billiard, a new mutual
separation is calculated ($\delta\theta_f$) and the separation ratio is
obtained ($p_a=|\delta\theta_f|/|\delta\theta_i|$). The initial separation is
renormalized by $\delta\theta_i=\delta\theta_f/p_a$. 
The method includes a real parameter $\mu$, which is positive if we seek unstable 
dynamics, and negative otherwise. 
If $p^{\mu}_a>1$ a clone of the trajectory and perturbation is created 
with a probability $p^{\mu}_a-1$, otherwise the trajectory is
killed with a probability $1-p^{\mu}_a$. For this method, the maximum number
of iterations considered are $10^{3}$. 

Figure \ref{Fig6}(a,b) shows a plot of the phase space (black dots) for 
different values of $g$. A period doubling bifurcation is observed for
$E=0.5$, $p=2$ and $\epsilon=0.1$ (oval billiard). For $g=0.1$ (Fig.
\ref{Fig6}(a)) one sees the green dots ($\mu=+1$) are circulating the
two large KAM islands, centering a saddle fixed point. The red dots
concentrate more inside the KAM islands. Then we use $\mu=-1$ in the LWD to
highlight the periodic islands. After increasing $g$ to $g=0.25$ (Fig.
\ref{Fig6}(b)) we see the green dots do not have a preferred region and
it is clear that a period doubling bifurcation happened and the red dots
follow the four big KAM islands. 

Now we use LWD to show for some special combinations of parameters that KAM
islands are not observed in the phase space, leading to an apparent ergodic
dynamics. For the simulations we consider $g=0.5$ and in Figs.
\ref{lwd_compara}(a,b,c,d) we have $\epsilon=0$ (circle billiard), in Figs.
\ref{lwd_compara}(e,f,g,h) we considered $p=2$ and $\epsilon=0.1$ (oval
billiard) and finally in Figs. \ref{lwd_compara}(i,j,k,l) we have $a=1.2$ and
$b=1$ (elliptic billiard). For $E=0.524$ in the circle (Fig.
\ref{lwd_compara}(a)), $E=0.602$ in the oval (Fig.\ref{lwd_compara}(e)) or
$E=0.667$ in the ellipse (Fig.\ref{lwd_compara}(i)) we observe that the
trajectories and clones for the LWD are highlighting a period four KAM island. 
Increasing slightly more the value of $E$ to $0.58$ for the circle (Fig.
\ref{lwd_compara}(b)), to $0.65$ for the oval (Fig. \ref{lwd_compara}(f)) and
to $0.71$ for the ellipse (Fig. \ref{lwd_compara}(j)) we see that the trajectories/clones
are not tending to go any specific region. Possibly these regions are ergodic,
but we could not prove the opposite, where several simulations were made
trying to identify small KAM islands. Considering the circle with $E=0.603$
(Fig. \ref{lwd_compara}(c)), the oval billiard with $E=0.675$ (Fig.
\ref{lwd_compara}(g)) and the elliptic billiard with $E=0.737$ (Fig.
\ref{lwd_compara}(g)), we see eight KAM islands. The behaviour of LWD for the
billiards are not the same, and they have more complex structures with
different periods that have not been shown. They have however similarities
near the seemingly ergodic region. Using $E=0.8$, $E=0.73$ and $E=0.9$,
respectively for the circle, oval and ellipse, we observe again the
trajectories/clones are not tending to go to any specific region.

In Figs. \ref{Fig8new_2}(a,b,c) it is shown $\theta$ as a function of $E$ for
each trajectory/clone in the circle, oval and elliptic billiard, respectively. The
control parameters are labeled in the figures. For a constant value of $E$ the
periodic regions can be identified as the ones to which $\theta$ does not vary
for long range.

Now we split both $\theta$ and $\alpha$ coordinates of the phase space in
$n_{div}$ different equally spaced regions. Therefore we have a grid with
different rectangles than can be measured if are occupied by a trajectory/clone in
the LWD. $f$ corresponds to the fraction of the number of rectangles occupied
over $n_{div}^2$. We consider in simulations $n_{div}=1000$. Figs.
\ref{Fig8new_2}(d,e,f) show $f~\rm{vs}~E$ for the circle, oval and elliptic
billiards. We see that the apparent ergodic region has $f\cong 0.02$ for the 
circle and $f\cong 0.05$ for the oval and ellipse. The regions with periodic
orbits tend to have $f\rightarrow0$. We see for the circle (Fig.
\ref{Fig8new_2}(d)) the ergodic and non-ergodic behaviors appear many times,
as indicated by the values of E in which $f<0.01$ for example. Therefore,
figures of $f~\rm{vs}~E$ in other opportunities can be used as an indicative
of ergodicity in a system.

\subsection{Supplemental Data II}

As a complementary material, we have as attached files a video showing the LWD
for $\mu=-1$ and considering $E\in[0.602,0.79]$ for the circle. The Figs.
\ref{lwd_compara}(a,b,c,d) show some frames of this video. The video can be
seen seen in the following electronic
address \href{http://www.youtube.com/watch?v=_V0G39yVG2I&feature=share&list=UUG_8BL4kcVV1rYYR4Zyn9ew}{http://www.youtube.com/watch?v=_V0G39yVG2I&feature=share&list=UUG_8BL4kcVV1rYYR4Zyn9ew}. 

\section{Time-dependent billiards}
\label{sec6}

\begin{figure}[htb]
\includegraphics[width=1.0\linewidth]{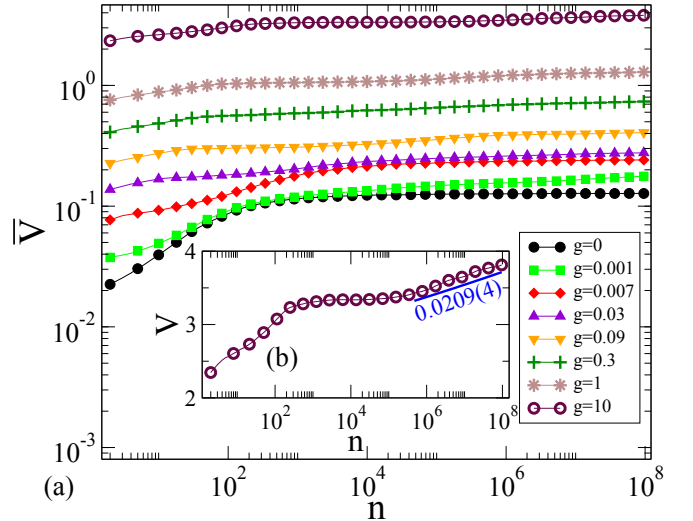}
\caption{(Colour online) For the circle billiard ($\epsilon=0$ and $a=b$), with 
$V_0=0.01$ and $\eta=0.01$ we have: (a) $\overline{V}~\rm{vs}~n$ considering different 
values of $g$; (b) Magnification near the curve with $g=10$, where it is possible to 
observe that the curve does not have enough time to saturate for $10^8$ iterations of 
the mapping.}
\label{Fige1}
\end{figure}

\begin{figure}[htb]
\includegraphics[width=1.0\linewidth]{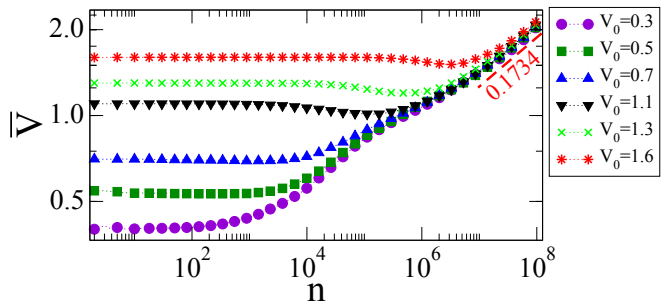}
\caption{(Colour online) For $\epsilon=0.2$, $p=2$, $g=0.1$ and $\eta=0.01$ we
have $\overline{V}~\rm{vs}~n$ considering different values of $V_0$.}
\label{Fige3}
\end{figure}

In this section we introduce a time-dependence in the boundary of the
oval/circle billiard. We consider that the boundary is breathing and the
radius is given by
\begin{equation}
R(p,\epsilon,\eta,\theta,t)=1+\eta \cos(t)+\epsilon[1+\eta\cos(t)]\cos(p\theta). 
\label{eqa1}
\end{equation}
For $\eta=0$ the static boundary case is recovered. Our main purpose is to
study and understand the influence of the gravitational field $g$ in the Fermi
acceleration phenomena. To do so, we define the average velocity as
\begin{equation}
\overline{V}(n)={{1}\over{M}}\sum_{j=1}^M\left[{1\over n}\sum_{i=1}^n
V_i\right], 
\end{equation}
where $M=500$ defines an ensemble of different initial conditions randomly
chosen in $\theta_0\in[0,2\pi)$, $\alpha_0\in[0,\pi)$ and $t_0\in[0,2\pi)$.

To discuss the results of the time dependent boundary, let us start with the
case of absent gravitation field, $g=0$. For the circle billiard, there is no
chaos in the phase space for the static version. Hence initializing the
simulation with a very low initial velocity, the curve of average velocity
starts to grow with and then suddenly changes to a regime of saturation. Such
regime marks a final limit of grow for the velocity of the particle then no
unlimited energy growth is observed. This is in well agreement to what is
known in the literature, particularly the LRA conjecture.

Then let us make $g\ne 0$ but considering small values, as labeled in the
figure. Because now $g\ne 0$, thin regions of the phase space for the
circular case have chaos for the static boundary. The curves of average
velocity seem not to saturate but rather they appear to grow at a very small
rate as shown in Fig. \ref{Fige1}(b) for a magnification considering $g=10$.
The slope of growth is $0.0209(4)$. Why this exponent is remarkably small?
Indeed according to Arnold \cite{arnold}, who proved there exist solutions
exhibiting arbitrarily large growth in the action variables in nearly
integrable dynamical systems with several degrees of freedom. Therefore we
believe few orbits are having diffusion in velocity while others are in
regular dynamics that, in the average are producing the behaviour shown in
Fig. \ref{Fige1}, as observed.

Now let us consider the oval billiard for $p=2$, $\epsilon=0.2$, $g=0.1$ and
$\eta=0.01$. It is shown in Fig. \ref{Fige3}(a) a plot of $\overline{V}$ as
function of $n$ for different initial velocities $V_0$. As one sees, the
higher $V_0$ the higher is the initial plateau $\overline{V}$ observed until
the curve changes to a regime of growth. The curves have an initial plateau
with constant velocity, and after passing by a crossover $n_x$, they start
growing with slope $\sim 0.1734$. We notice this result is slightly larger
than the one obtained \cite{ref19}, i.e. $\sim 0.16\ldots$ for the breathing
case and foreseen theoretically in \cite{robnik} as $\sim 1/6$. However our
result is within the same order of magnitude of either results obtained
previously.

\section{Conclusions}
\label{sec7}

We studied some classical particles undergoing collisions inside in a circle or 
oval billiard under a gravitational force field. The mapping and model were 
studied and details about the dynamic of these particles were realized. 
The linear stability of some fixed points was studied and it was possible 
to obtain analytically and numerically the conditions where we have a 
period-doubling bifurcation of a period one and two 
fixed points.  We discovered apparently ergodic dynamics for certain values
of the control parameters, which passed through a sensitive test using
Lyapunov weighted dynamics. 
After introducing a time-dependence in the boundary, we showed
for the circle with null gravity $g$ that the system has average velocity that tends 
to saturate after some number of iterations $n$, but for $g>0$ it is possible to see
that we have a Fermi acceleration phenomenon with slope very small, where
this phenomena can be explained using the LRA conjecture and Arnold diffusion.
We observed that the slope for the oval billiard 
is slightly greater than the one obtained in the Ref. \cite{ref19}, but before the 
Fermi acceleration the particles tend to enter in a deceleration region for high 
enough values of initial velocity.

The numerical finding of ergodicity (strengthened by Lyapunov-weighted dynamics)
for a smooth convex gravitational billiard is new; it would be interesting
to obtain rigorous mathematical proofs for the examples given here, and determine
conditions analagous to the defocusing mechanism for non-gravitational billiards
that are sufficient for ergodicity.   In addition, the dynamics of ergodic billiards
is largely determined by the amount ``stickiness,'' regular (albeit zero measure)
orbits that trap trajectories for long periods.  For example, the level of stickiness
in the chaotic region of mushroom billiards is controlled by the existence of
marginally unstable periodic orbits (MUPOs), which in turn is related to approximation
of real numbers by certain rationals~\cite{extra_1}.  It would be interesting to know if the
gravitational billiards considered here have dynamics controlled by similarly sticky orbits.

Likewise, we have observed Fermi acceleration in the gravitational circular billiard, despite
the fact that the high velocity limit is integrable.  While this is apparently due to Arnold
diffusion, it would be very interesting to explore this mechanism in more detail, and in
general the limit of high velocity of gravitational billiards with diverse geometries.  For
the oval billiard, the acceleration exponent is consistent with other (non-gravitational)
Fermi acceleration problems, but this need not be the case in general.  Finally, we note
that both gravitational effects and time-dependent boundaries can be explored through
atom-optics experiments~\cite{ref12}.

\section*{Acknowledgements}

DRC acknowledges Brazilian agency FAPESP. EDL thanks to CNPq, FUNDUNESP and
APESP, Brazilian agencies. This research was supported by resources supplied
by the Center for Scientific Computing (NCC/GridUNESP) of the S\~ao Paulo
State University (UNESP). This work was finished during a visit of DRC as a
PhD sandwich supported by FAPESP to the School of Mathematics from the
University of Bristol.

\end{document}